\documentstyle[12pt,psfig,aaspp4]{article}

\lefthead{Rubenstein \& Bailyn}
\righthead{Observations of the core of NGC 6752}

\begin{document}

\title{HST Observations of the Central-Cusp Globular Cluster NGC 6752.
The Effect of Binary Stars on the Luminosity Function in the Core}

\author{Eric P. Rubenstein \& Charles D. Bailyn}
\affil{Yale University Astronomy Department \\
PO Box 208101, New Haven, CT, 06520-8101}
\authoraddr{PO Box 208101, New Haven, CT, 06520-8101\\
ericr@astro.yale.edu\\
Tel \# (203) 432-3028\\
FAX \# (203) 432-5048}

\begin{abstract}

We consider the effect of binary stars on the main-sequence luminosity
functions observed in the core of globular clusters, with specific
reference to NGC~6752.  We find that mass segregation results in an
increased binary fraction at fainter magnitudes along the
main-sequence.  If this effect is not taken into account when
analyzing luminosity functions, erroneous conclusions can be drawn
regarding the distribution of single stars, and the dynamical state of
the cluster.  In the core of NGC~6752, our HST data reveal a flat
luminosity function, in agreement with previous results.  However,
when we correct for the increasing binary fraction at faint
magnitudes, the LF begins to fall immediately below the turn-off.
This effect appears to be confined to the inner core radius of the
cluster.

\end{abstract}

Subject headings: binaries: general --- globular clusters: general ---
globular clusters: individual (NGC 6752) --- stars: luminosity
function --- stars: population II

\section{Introduction}

The study of a globular cluster's luminosity function (LF) provides
insight into its present dynamical state and the stellar populations
of which it is comprised.  However, the presence of binary stars can
alter the appearance of the luminosity function.  A LF constructed
from a population containing a significant fraction of binary stars is
not a single star LF at all, but rather an amalgam of single stars and
binaries.  Because of mass segregation, the binary fraction in a
cluster is likely to vary with magnitude and radial distance to the
cluster center.  Thus we expect in general that the presence of
binaries may make the single star LF different from the observed
main-sequence LF, which includes stars on the binary sequence.

Here we attempt to quantify this effect using our HST data of
NGC~6752, in which we have previously discovered a large, centrally
concentrated population of main-sequence binary stars in the core
(Rubenstein \& Bailyn, 1997 hereafter Paper~II).  Data reduction
and calibration procedures are discussed in Paper~II.  Here we
discuss the implication of the binary sequence we discovered
for the cluster LF.

\section{Determining The Luminosity Function and the Effects of Mass
Segregation}\label{LF}

To disentangle the true LF from an uncorrected LF it is necessary to
perform artificial star tests.  In Paper~II we describe the procedure
we employed to digitally add nearly $10^7$ artificial stars to the
images.  We demonstrated that the artificial stars had photometric
errors which were very similar to the real stars, and should therefore
have the same recovery probabilities as real stars.  The artificial
stars were added with a flat LF which was similar to the observed LF.
We calculated the recovery rate of artificial stars in a fashion
similar to Bolte (1994), although we used magnitude bins of 0.5 mag.
Briefly, the fraction of artificial stars recovered in the $i^{th}$
magnitude bin, $f_i$, is the number of stars recovered within that
magnitude bin, divided by the number of stars added to the data in
that magnitude bin.

We calculated the incompleteness correction factor by inverting the
recovery fraction, 1/$f_i$.  Since we have only split the data into
``inner'' and ``outer'' regions, we did not construct a
two-dimensional completeness look-up table as did Bolte.  Rather, we
separately calculated the incompleteness corrections for each region.
We then smoothed the results by performing a least-squares fit to the
recovery fraction values as a function of magnitude.
We did not fit the data beyond V=23 since at this point the
completeness drops suddenly by more than a factor of 2 to below 50\%.
Due to the large saturated regions near a few extremely bright stars,
where no objects are recovered at all, even relatively bright stars
are only about 75\% complete.  However, only the relative completeness
from magnitude bin to magnitude bin is relevant to a discussion of the
LF of the cluster stars.

\begin{figure}
\centerline{\psfig{file=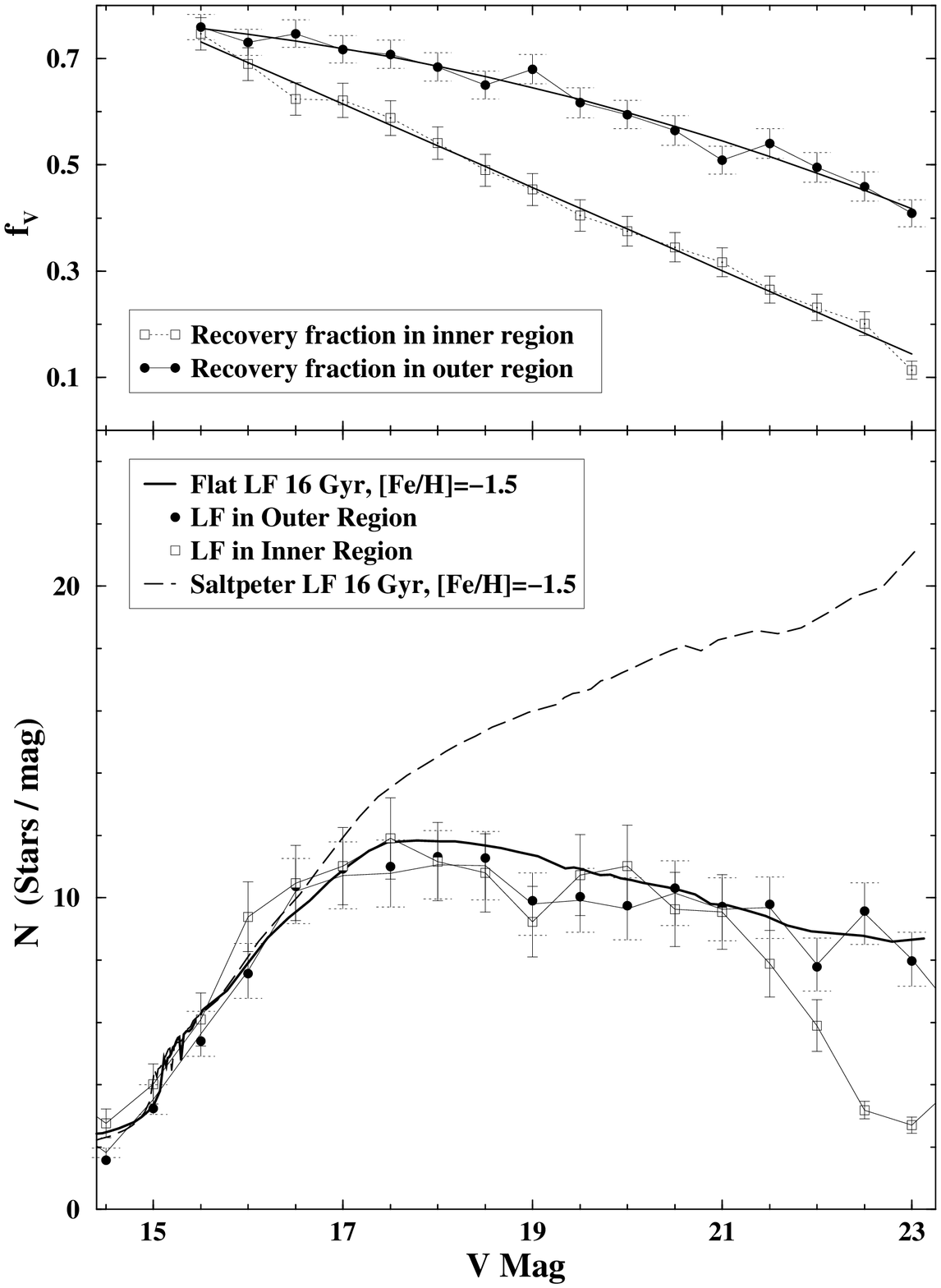,height=5.0in}}
\caption{Top- Fraction of recovered artificial stars for V images.
The error bars indicate the Poisson uncertainty of the recovery
fraction.  To perform the completeness corrections on the LF, we first
performed least-squares fits and then evaluated these functions at the
magnitude of each star before weighting each star by the interpolated
factor 1/f.  Least-square fits are not performed beyond the point at
which recovery fractions fall precipitously. \protect\\ Bottom- The
LFs for the instrumental V photometry in the inner (squares) and outer
(circles) regions with 2 theoretical LFs superposed.  An arbitrary
factor has been applied so that the observed LFs have the same values
along the sub-giant branch and at the MSTO.  Note that the LF is flat
for 5 mags below the turn-off in the inner region, and for 7 mags in
the outer region.  The theoretical LFs are taken from a 16 Gyr, Yale
Isochrone (Chaboyer et al. 1995) with [Fe/H]=-1.5 (Djorgovski 1993).
The LF which represents a flat IMF is plotted as a bold, solid line,
and the LF due to a Salpeter IMF is the dashed line.}
\label{LFpic}
\end{figure}

To construct the LFs, we bin the stars into 0.5 mag bins.  Each star
receives a weight equal to the inverse of its recovery probability.
The results, and our interpolation, are shown in Figure~\ref{LFpic}.
We have scaled the LF of the inner regions by an arbitrary factor so
that the inner and outer regions have the same values along the
sub-giant branch and at the main-sequence turn-off (MSTO).  In the
inner region ($\sim 1 r_{core} \simeq 12\arcsec = 0.2$pc; Djorgovski
1993), the V-band LF is flat for 5 magnitudes below the MSTO, which is
located at V$\sim16.5$.  Beyond 5 magnitudes below the MSTO, the LF
falls very rapidly.

Both the plateau and the sudden drop can be attributed to the advanced
dynamical state of this stellar population.  In particular, mass
segregation will eject the low mass stars from the cluster center and
force the more massive objects from the outer parts of the cluster in
towards the center (see review by Heggie \& Meylan 1997).  The
``outer'' region retains more of its low mass objects, in that it has
a flat LF 2 magnitudes further down from the MSTO than the inner
region.  Converting from luminosity to mass using the Yale Isochrones
(Chaboyer et al. 1995), it is clear that low and moderate mass objects
are strongly depleted relative to a Salpeter IMF.  Indeed, the inner
region shows an inverted mass function beyond 5 magnitudes below the
furnoff.  This result is statistically significant, but caution is
warranted because the completeness level has dropped by about a factor
of three.

\section{Binary Fraction as a Function of Luminosity}\label{binseg}

In general, one would expect that mass segregation in a GC core would
result in a larger binary fraction of the lower main sequence than
near the turnoff, because low mass main sequence stars would be
preferentially ejected from the core.  To test this hypothesis, we
checked to see if the color distribution redward of the main-sequence
ridge-line (MSRL, as defined in Paper~II) was a function of magnitude.
We split the magnitude interval $16.5\leq$V$\leq 19.0$ into two equal
portions, $16.5\leq$V$\leq 17.75$ (hereafter the ``bright'' stars) and
$17.75 \leq$ V $\leq 19.0$ (hereafter the ``dim'' stars).  Then we
performed an analysis identical to the one described in \S~3.2 of
Paper~II.  Briefly, we performed Monte Carlo experiments using the
photometric results of both the real stars and the artificial stars.
We calculated the difference in color, $\Delta$C, between the MSRL and
each star (both real and artificial).  We then determined a parameter
$Y$ for each real star, equal to the fraction of artificial stars of
similar magnitude and crowding which have $\Delta$C smaller than that
of the real star.  If the real stars and artifical stars are drawn
from the same input distribution, the values of $Y$ should be evenly
distributed from zero to unity.  The fact that they were not
demonstrated the need for an underlying population of binary stars
(see Paper II for more details).  We found that the distribution of
$Y$ values was significantly further from uniform among the dim stars
than among the bright stars (Figure 2), indicating a greater fraction
of binary stars in the dim group, as expected.  The formal probability
that the two groups of stars were drawn from a population having the
same input distribution was $10^{-7}$.

To quantify the variation of binary fraction with magnitude, one would
ideally determine the absolute binary fraction among both the bright
and dim star populations.  Unfortunately, by splitting the stars into
two groups the statistical significance of the results are degraded.
Specifically, the 3$\sigma$ limits of the binary fraction determined
in the manner of Paper II are poorly constrained, 4\%---50\% for the
bright stars, 18\%---42\% among the dim stars.  Fortunately, it is
possible to obtain statistically significant results by calculating
the {\it difference} in binary fraction between the bright and faint
stars.  To perform this differential analysis, we modified
the Monte Carlo procedure discussed in Paper~II for determining the
binary fraction.  In this case, we increased the magnitude of some of
the stars in the bright group to simulate the effect of additional
binaries.  We then compared the $Y$ distribution of this altered
bright star population with that of the dim stars --- the fraction of
bright stars which had to have light added for the two distributions
to be comparable is a measure of the difference in the binary star
population of the two groups.

In carrying out this procedure, we had to specify the ratio of
brightness of the two stars in our fake binary systems. Since the
distribution of binary mass ratios and luminosity ratios is unknown,
we used the equation $V_2=\frac{ V_1} {R^\xi },$ where $R$ is a random
number between 0 and 1.  This relation, while convenient to work with,
is not meant to accurately model what is, after all, an unknown
distribution.  The free parameter $\xi $ determines the luminosity
ratio of the binary distribution: $\xi=0$ corresponds to the case
where all binaries have components with equal luminosity, while larger
values of $\xi $ result in distributions increasingly weighted toward
smaller luminosity ratios (and thus small mass ratios).

\begin{figure}
\centerline{\psfig{file=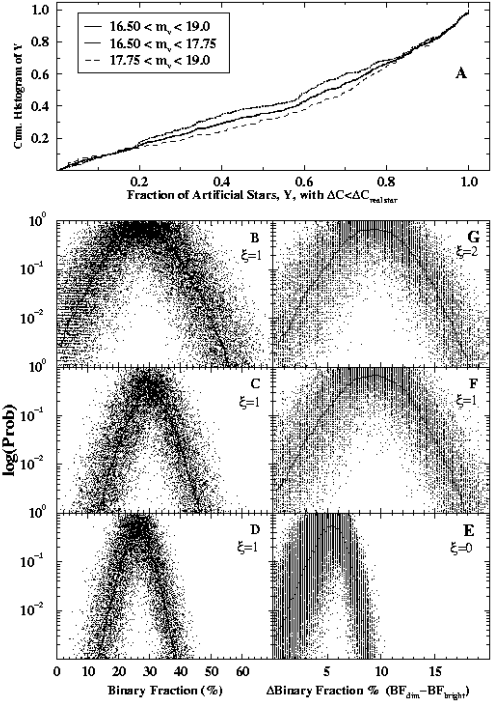,height=4.0in}}
\caption{A- The $\Delta$C distributions (see \S\ref{binseg}) of the
``bright'' and ``dim'' populations of stars (top and bottom lines,
respectively) are different from each other and from the total
population (middle line).  The formal probability that the bright and
dim stars are drawn from a population which has a single $\Delta$C
distribution is $10^{-7}$.\protect\\ 
B,C \& D- Panels B-D show the required binary fraction among the stars
for the case $\xi=1$ (see \S\ref{binseg}).  ``B'' shows the results
for the bright stars alone, the ``C'' displays the dim stars' results,
and ``D'' represents the case where all of the stars are considered
together.  Due to statistical uncertainties, it is impossible to
measure the absolute binary fraction of the bright and dim stars
separately.\protect\\
E, F \& G- Panels E-G show the {\it difference} in binary fraction
between the bright and dim stars.  ``E'' indicates that under the
assumption that all binaries have equal mass components, the dim group
of stars must have about 5\% more binaries than the bright group.
Panels ``F'' \& ``G'' show the results for non-equal mass binaries,
``F'' for the case $\xi=1$ (see \S\ref{binseg}), and ``G'' for
$\xi=2$. In both of these cases the dim stars have about 10\% more
binaries than the bright stars.}
\label{bey}
\end{figure}

Figure~\ref{bey} shows the results of these calculations.  We found
that for $\xi=0$, $\sim 5$\% of the bright stars must have binary
companions added to match the dim stars' distribution.  For the
physically more realistic cases $\xi=$1 or 2, $\sim 10$\% of the
bright stars must ``become binaries'' to match the distribution of the
dim stars.

\section{Correcting the Luminosity Function for Binaries}\label{binlf}

The relative LF of single stars would be minimally altered if the
binary fraction were constant at all magnitudes.  If that were the
case, there would be equal numbers of binaries in all LF bins, and the
underlying single star LF would not be masked.  However, in NGC~6752
we now know that the binary fraction (BF) does change with magnitude.
Therefore, the single star LF {\em is} altered by the binary
population and can not be observed unless we first account for the
binaries.  In this section we will give an example of how to perform
this correction.  This calculation is not intended to be definitive,
but rather to demonstrate the potential size of the effect on the LF.

Since the binary frequency changes with magnitude, but is only defined
in two magnitude ranges, we must make assumptions about the way the
binary population changes.  To minimize the number of free parameters,
we will assume that the BF varies linearly with magnitude.  To
determine the BF in each magnitude bin, we combine the absolute binary
fraction (from the results of Paper~II) with a simple, linear
extension of the magnitude dependence of the BF (found in above).  The
average magnitude of the stars analyzed in Paper~II, in the interval
from V=16.5 to V=19.0, is 17.75.  We assume that at this
representative magnitude, the binary fraction is the mean of the
$3\sigma$ limits derived in Paper~II, ie. $(0.15+0.38)/2=0.265$.  We
determined above that the binary fraction increases 10\% from the
bright stars to the dim stars.  Using the midpoints of those groups'
magnitudes we determine that at V=17.125, the binary fraction is
$0.215$ and that every 0.5 mag fainter on the CMD corresponds to an
increase of 4\% in the binary fraction.

\begin{figure}
\centerline{\psfig{file=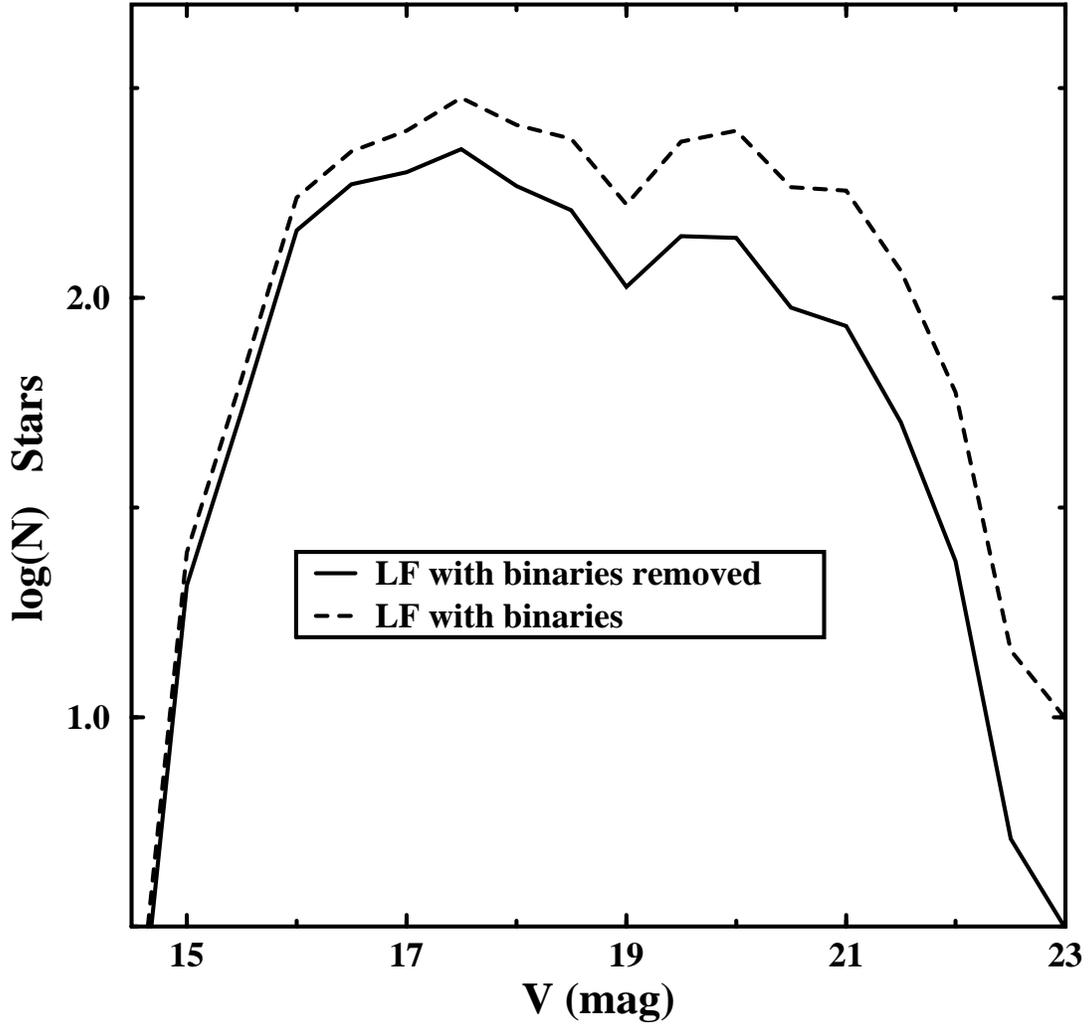,width=5.6in}}
\caption{Removing binaries from the star counts further depresses the
LF, actually inverting it almost all the way to the turn-off at
V$\approx 16.5$ mag.}\label{lfnobinaries}
\end{figure}

We then construct a new LF in which the number of stars in each
magnitude bin is reduced by an amount equal to the binary fraction.
In other words, for a given magnitude bin the ``binary fraction
corrected star count'', ${\rm N}_b = {\rm N}_c{\rm *(1-BF), } $ where
N$_c$ is the completeness corrected star count and BF is the binary
fraction in that magnitude bin.  In Figure~\ref{lfnobinaries} we plot
the resulting V-band LF from the inner region, along with the version not
corrected by the removal of binaries.  Not surprisingly, the LF is
even more depressed at faint magnitudes when the effect of binaries is
considered.  The inversion noticed at fainter magnitudes in the LF of
the core, plotted in Figure~\ref{LFpic}, is now present virtually all
the way to the turn-off region.  This calculation shows that
neglecting binaries will lead to qualitative errors in the derived LF.
However, the assumptions required about changes in binary fraction
with magnitude mean that these particular results may not be
quantitatively reliable.

In Paper~II we speculated that at faint magnitudes the MSRL might be
shifted significantly to the red due to a preponderance of binaries at
the faint end dominating over low luminosity single stars.  The
results obtained here support an interpretation that the observed
ridge-line is not the {\em main-sequence} ridge-line below about 3.5
magnitudes below the MSTO, but rather, at fainter and fainter
magnitudes, it is increasingly a {\em binary} ridge-line.  Under our
crude assumptions, the BF at V=21 is over 50\%.

Ground based studies by Da Costa (1982) and Richer et al. (1991) found that
the LF in NGC~6752 away from the cores rises all the way to the
faint object cutoff at $m_v=22.5$ and $m_v=23.5$ respectively.
Recently, Shara et al. (1995) and Ferraro et al. (1997) have shown
from HST data that the LF flattens closer to the core.  Both our
inner LF uncorrected for binaries, and our outer LF are in
general agreement with the flat LF found in the core by Shara et al. 
(1995).  However, the inverted single star LF suggested for the
core by the binary correction described here implies a greater
degree of dynamical evolution than does the previous work.

\section{Conclusions}

Our analysis of the luminosity function in the core of this cluster
indicates that for about 5 magnitudes down from the main-sequence
turn-off there is a flat LF and beyond that point the LF is falling.
Below this point there is an inverted mass function.  However, this LF
does not represent the LF of single main-sequence stars because there
are more binaries at fainter magnitudes, presumably due to mass
segregation.  We find that the population of binaries increases by
about 8\% per magnitude over the small interval we were able to test.
When we extrapolate this trend to fainter magnitudes, which may not be
justified, we find that for single stars there is evidence of an
inverted mass function nearly all the way up to the MSTO.  Another
implication of this extrapolation is that below about 3.5 magnitudes
below the MSTO the observed ridge-line is dominated by binaries to the
extent that it is significantly different from the single star
ridge-line, and should therefore not be used for isochrone fitting.

Future studies of the stellar populations of GC cores must account for
the effect of binaries.  Failure to correct for the binaries on the
lower main sequence will have the effect of over estimating the number
of low mass stars.  Since the binaries are located preferentially in
the core it is unlikely that they will alter the results in the outer
regions of clusters.

\section{Acknowledgments}

EPR would like to thank Peter Stetson, Ken Janes and Jim Heasley for
making newer versions available of DAOFIND and SPS.  CDB is grateful
for a National Young Investigator award from the NSF.  We thank
Adrienne Cool, Pierre Demarque, Richard Larson, Mario Mateo, Jerry
Orosz \& Alison Sills for comments and suggestions.  Sukyung Yi
provided detailed instructions on how to transform Yale Isochrones to
HST WFPC2 filters (detailed in Yi, Demarque \& Oemler 1995), and
Alison Sills carried out this transformation.  Mary-Katherine McGovern
assisted with the artificial star tests.  This research has made use
of the Simbad database, operated at CDS, Strasbourg, France.  This
work has been supported by NASA through LTSA grants NAGW-2469 \&
NAG5-6404 and grant number HST-GO-5318 from the Space Telescope
Science Institute, which is operated by the Association of
Universities for Research in Astronomy, Inc., under NASA contract
NAS5-26555.

\eject

\end{document}